# Lattice-decoupled rotatable stripe-like charge order within the strange metal phase of 2M-WS$_2$

Kebin Xiao[a,b,1], Yunkai Guo[a,b,1], Daran Fu[a,b], Yuqiang Fang[c], Yating Hu[d], Jingming Yan[a,b], Yucong Peng[a,b], Yuyang Wang[a,b], Yongkang Ju[e], Peizhe Tang[e,f], Xiangang Wan[d], Fuqiang Huang[c,g,2], Qi-Kun Xue[a,b,h,i,j,2], and Wei Li[a,b,j,2]

[a]State Key Laboratory of Low-Dimensional Quantum Physics, Department of Physics, Tsinghua University, Beijing 100084, China; [b]Frontier Science Center for Quantum Information, Beijing 100084, China; [c]School of Materials Science and Engineering, Shanghai Jiao Tong University, Shanghai 200240, China; [d]National Laboratory of Solid State Microstructures and School of Physics, Nanjing University, Nanjing 210093, China; [e]School of Materials Science and Engineering, Beihang University, Beijing 100191, China; [f]Max Planck Institute for the Structure and Dynamics of Matter, Center for Free Electron Laser Science, 22761 Hamburg, Germany; [g]State Key Laboratory of High Performance Ceramics and Superfine Microstructure, Shanghai Institute of Ceramics, Chinese Academy of Science, Shanghai 200050, China; [h]Beijing Academy of Quantum Information Sciences, Beijing 100193, China; [i]Southern University of Science and Technology, Shenzhen 518055, China; [j]Hefei National Laboratory, Hefei 230088, China

[1]K.X. and Y.G. contributed equally to this work.

[2]To whom correspondence may be addressed. Email: huangfq@sjtu.edu.cn, qkxue@mail.tsinghua.edu.cn, or weili83@tsinghua.edu.cn.

**This PDF file includes:**

    Main Text
    Figures 1 to 5

**Abstract**

In quantum materials, charge orders typically stabilize in specific crystallographic orientations, though their formation mechanisms may vary. Here, using low-temperature scanning tunneling microscopy (STM), we discover a lattice-decoupled rotatable stripe-like charge order coexisting with superconductivity in 2M-$WS_2$. The charge order manifests five distinct orientations across different sample regions, yet maintains an identical wavelength. This directional decoupling from host lattice challenges existing paradigms. First-principles calculations of phonon spectra and nesting function fail to explain the ordering mechanism. Intriguingly, the transition temperature of the charge orders exhibits spatial variations (21-46 K), coinciding with the temperature range of the recently reported strange metal phase in this material. This correlation suggests that the interplay between strong electronic correlations and electron-phonon coupling must be critically evaluated to elucidate the emergence of this unconventional charge order.

**Significance Statement**

We report a paradigm-breaking charge order in 2M-$WS_2$ that decouples from lattice symmetry, exhibiting fivefold rotational freedom with uniform periodicity. This order coexists with superconductivity while its transition temperatures reside within the material's strange metal phase, bridging the two enigmatic phenomena in strongly correlated systems, and echoing debates on phase competition in cuprates. First-principles calculations rule out conventional nesting and phonon mechanisms, demanding new models incorporating strong correlations. The lattice decoupling and strange metal correlation collectively challenge the theoretical separation between electronic and lattice degrees of freedom, necessitating unified frameworks that bridge strong correlations, phonon interactions, and non-Fermi liquid physics.

**Main Text**

**Introduction**

Charge orders are extensively observed across quantum materials. In weakly correlated systems, charge order predominantly arises from electronic instabilities driven by Fermi surface nesting or momentum-selective electron-phonon coupling (1-3). In contrast, in strongly correlated systems, enhanced electronic interactions involving intertwined charge, spin, and orbital degrees of freedom emerge as dominant governing mechanisms (3-13).

Transition metal dichalcogenides (TMDCs) exhibit intriguing electronic properties, such as superconductivity, charge order and electronic topology (14-24). A noteworthy material within the TMDC family is 2M-$WS_2$, distinguishing itself with the coexistence of $Z_2$ non-trivial topological properties and bulk superconductivity ($T_c$~8.8 K) (21-34). Moreover, strong electron-phonon coupling (33) and anomalous enhancement of the Nernst effect (31) have been observed in 2M-$WS_2$, indicating the existence of a possible strange-metal state above the Fermi-liquid state as well as the superconducting state (31). A surface charge order has also been observed, and the interplay between the charge order and the electronic topology has been studied (24).

In this article, we report the discovery of a paradigm-breaking rotatable stripe-like charge order in 2M-$WS_2$ via scanning tunneling microscopy (STM). The charge order exhibits five distinct orientations with identical periodicities across different regions of samples. The charge order's transition temperatures (21-46 K) reside entirely within the recently reported strange metal phase (31), indicating strong electronic correlations act as the dominant driver stabilizing this ordered state.

## Results
### Rotatable stripe-like charge order

2M-WS$_2$ forms a van der Waals crystal through the stacking of 1T'-WS$_2$ monolayers along the *c*-axis. Within each monolayer, W atoms deviate from the center of [WS$_6$]$^{8-}$-octahedra, resulting in a zigzag structure along the *a*-direction (Fig. 1 *A-C*) (21,22). Besides the zigzag atomic chains (denoted by green arrows in Fig. 1 *D-G*), we also observe long-range stripe-like electronic modulations (charge order) on WS$_2$ in our STM topographic images (Fig. 1 *D-G*). The orientations of stripe-like charge orders are denoted by yellow arrows, and notably, the angles between the directions of stripes and the zigzag structures at different locations and in different samples are variable (Fig. 1 *D-G* and *SI Appendix*, Fig. S1). The measured angles can be 45°, 51°, 72°, 80° and 90°. In the regions with smaller angles, the stripe orientation appears more uniform, resulting in a pair of sharp point-like $q_{stripe}$ signals in the Fourier transform (FFT) images (insets of Fig. 1 *D2-F2*). While, in the regions where the angles are close to 90°, the stripes exhibit irregular orientations, corresponding to line-like FFT signals (inset of Fig. 1*G2*). By extracting the corresponding signal $q_{stripe}$ from the FFT and applying an inverse Fourier transform (IFFT), we confirm that the regions with smaller angles display consistent unidirectional stripe-like patterns (Fig. 1 *D2-F2*). In contrast, the region with the 90° angle shows distorted stripe-like patterns (Fig. 1*G2*) (24). Note that the wide stripe-shaped modulations observed in STM topography (Fig. 1 *D1* and *E1*, see details in the later discussions), spanning multiple zig-zag chains along their axis and superimposed on stripe charge orders, originate from strain generated during sample cleaving.

Surprisingly, although the orientation of the stripes varies across different locations, the length of the wavevector $q_{stripe}$ remains identical. A statistical analysis of the lengths of the $q_{stripe}$ and the stripe orientations across all observed regions reveals that the periodicity of the stripes is consistently around 1.22 nm (Fig. 1*H*). Bias-dependent d*I*/d*V* mapping reveals the stripes persist across -20 to 32 meV with invariable periodicity, unambiguously confirming their static ordering nature (see details in *SI Appendix*, Fig. S2).

### Thermodynamic evolution: transition temperature and relaxation

To determine the transition temperature of the charge order ($T^*$), we performed temperature-dependent measurements. As the temperature increases, the strength of the stripes gradually diminishes until it disappears (upper panel of Fig. 2*A*). This trend is also evident in the FFT of the topography images (lower panel of Fig. 2*A*), where an attenuated $q_{stripe}$ signal is observed with a rise of temperature. By extracting the $q_{stripe}$ values and performing an IFFT at various temperatures, we clearly demonstrate that the stripes vanish at elevated temperatures (Fig. 2*B*). We plot the intensity of the $q_{stripe}$ as a function of temperature, and compile the results into a phase diagram (Fig. 2*C*) (31). The determined $T^*$ here is around 21 K. Actually, for different locations and samples, $T^*$ fluctuates from 21 K to 46 K (see details in *SI Appendix*, Note 3). These values precisely overlap with the reported transition temperature range of a strange metal phase (31) of 2M-WS$_2$, suggesting a potential connection between the stripe order and the strange metal phase (see details in the later discussions).

Bi-directional stripe charge orders are also observed in certain region, and the stripes tend to be rearranged into a uniform direction in our temperature-dependent experiments. Within the region, two different orientations of stripe exhibit mirror-symmetric with respect to the direction of the zigzag chains (Fig. 3*A*). When the sample temperature was raised above $T^*$, and subsequently cooled back down to 4 K, the initially bi-directional stripes become aligned. In subsequent annealing cycles, including multiple heating and cooling runs, the orientation of the stripes remains unchanged. In the FFT of the topographic images, before the first annealing cycle, both $q_{stripe}$ and $q'_{stripe}$ exist, corresponding to the two directions of the stripes (left panel of Fig. 3*B*). As the temperature exceeds *T*\*, both signals vanish simultaneously (middle panel of Fig. 3*B*). Upon further cooling to 4 K, only the $q_{stripe}$ signal reappears (right panel of Fig. 3*B*). We plot the intensity of $q_{stripe}$ and $q'_{stripe}$ as a function of the times of annealing cycle (Fig. 3*C*), which clearly shows that only $q_{stripe}$ survive after the cycles. During annealing cycles, the stripe-zigzag coupled modulation also disappears above *T*\* and recovers reversibly below *T*\* (see details in *SI Appendix*, Note 4).

**Interplay between charge order, superconductivity and magnetic fields**

We now investigate how the stripe-like charge order affects the superconductivity. The stripes coexist with superconductivity. A series of d*I*/d*V* spectra taken across the stripes Fig. 4*A*) reveal that the superconducting coherence peaks are modulated by the charge order. We obtain the spatial evolution of the peak heights and compare it with the spatial modulations of the stripes Fig. 4*B*). The results indicate that they are synchronous and share the same period. Applying an out-of-plane magnetic field can lead to the formation of vortices, which can host Majorana bound states on the surface of 2M-WS$_2$ (21,24). By cycling the magnetic field from zero to higher values and back down, we observe changes in the spatial distribution of the vortices (Figs. 4 *C1-C4* and *SI Appendix*, Fig. S8), indicating the weak pinning effect of the stripes on magnetic vortices. The vertical magnetic field does not influence the spatial arrangement of the stripes, as its distribution remains consistent across varying the fields up to 12 T (*SI Appendix*, Fig. S9).

**Spectrum signatures of charge order**

Next, we show the d*I*/d*V* spectrum signatures of the regions with the stripe-like charge order. Figure 5*A* shows an area, where stripe and no-stripe regions coexist. The spatial distribution of the amplitude of the stripes clearly shows the two distinct regions (Fig. 5*B*). The d*I*/d*V* spectrum taken on the no-stripe region exhibits a higher density of states (denoted by yellow shade in Fig. 5*C*) near the Fermi level ($E_F$), while the spectrum taken on the stripe region exhibits a dip-like feature near $E_F$ (denoted by gray shade in Fig. 5*C*). Remarkably, a broad peak-like feature around 100 meV can be observed in the spectra of the stripe region (denoted by red shade in Fig. 5*C*). Those phenomena are more obviously shown in a series of spectra taken along a line cut across the two regions (Fig. 5*D*). The peak-like feature at 100 meV has been confirmed in multiple samples. We compare the spectra taken below and above *T*\*, and the density of states near 100 meV in the two spectra are not changed (Fig. 5*E*). Consequently, the hump-like feature at 100 meV serves as a spectral characteristic signature of the regions predisposed to stripe order formation, although the associated electronic states, located far from $E_F$, do not actively participate in the phase transition. The connection between the stripes and the feature at 100 meV warrants further investigation.

**Discussion**

Lastly, we would like to discuss the origin of the observed rotatable stripe-like charge order. Since the calculated and measured band structures are in good agreement with each other (27-29) and the stripe-like charge order is incommensurate, we first try to understand the stripes under the Fermi surface nesting picture. We perform the nesting function calculation, and fail to reproduce the electronic instability at the observed $q_{stripe}$ with five different orientations (See details in *SI Appendix*, Note 8 and Note 9). We do not observe any negative phonon mode or phonon softening in the calculated phonon bands either (22,24). Furthermore, simulations unequivocally exclude the possibility of stripe modulations as moiré patterns (see details in *SI Appendix*, Note 10). Moreover, local strains may modify the electronic structures of a material, potentially influencing charge order or the strange metal phase. During the sample cleaving of 2M-WS$_2$, local strains can indeed be introduced, manifesting as the wide stripe-shaped modulations superposed onto the stripe charge orders in STM topography (Fig. 1 *D1*, *E1* and Fig. 2*A*). However, in certain areas where no observable strains are present, stripe charge orders still appear, for instance, in Fig. 1 *F1* and *G1*. This finding suggests that strain may not be necessary for the formation of the lattice-decoupled charge order. In addition, the identical wavelength across rotatable charge orders stands in stark contrast to the randomly distributed uncontrolled strain. Bias-dependent d*I*/d*V* mapping shows the stripe charge order persist across -20 to 32 meV with invariant periodicity (*SI Appendix*, Fig. S2), inconsistent with structural modulations detectable at high bias. Therefore, the conserved wavelength, energy persistence, and presence in strain-free zones demonstrate the electronic origin of the stripe charge order.

Notably, the orientations of the observed stripes are decoupled from the underlying lattice, maintaining identical periodicity while rotating freely relative to the lattice. Crucially, this invariance in the length of $q_{stripe}$ across all regions eliminates scenarios where modulations arise from periodic distortions localized to individual zigzag chains. In such cases, the projection of $q_{stripe}$ along the zigzag direction would remain fixed—a prediction incompatible with our observations of orientation-dependent projections (Fig. 1*H*). This rotational decoupling constitutes a previously unidentified class of charge order, likely arising from the unique electronic properties of 2M-WS$_2$. Intriguingly, this material displays: 1) a strange metal phase (20-120 K) evidenced by anomalous Nernst effect enhancement (31), and 2) dramatically strengthened electron-phonon coupling below ~100 K revealed through Raman spectroscopy (33). Given that $T^*$ of the stripe-like charge order fluctuate from 21 K to 46 K, embedded within the lower thermal boundary of the strange metal regime, we suspect that the electronic interactions and electron-phonon coupling, which are strongly strengthened, cooperatively drive the unconventional ordering. The discovered intimate connection between the charge order and the strange metal phase in WS$_2$ echoes the debates on phase competition in cuprate (35) while offering new material-specific insights. The lattice decoupling demonstrates that electronic self-organization can occur independently of structural constraints, while the correlation with strange metal implicates quantum fluctuations as a potential ordering trigger (36). This calls for a revision of current paradigms of charge order to incorporate correlated quantum fluid behaviors alongside traditional lattice-mediated mechanisms.

**Materials and Methods**

**Sample preparation**

The synthesis of 2M-WS$_2$ single crystals is initiated by the preparation of K$_{0.7}$WS$_2$, which is accomplished through the amalgamation of W, S, and K$_2$S$_2$ powders in a stoichiometric ratio within an argon-filled glovebox. The mixture was pressed into a pellet, and sealed in an evacuated silica tube at $10^{-5}$ Torr. The tube was heated to 850°C at a rate of 5°C/min, maintained at this temperature for 6000 minutes, and then cooled to 600°C at a rate of 0.1°C/min in the muffle tube. The as-synthesized K$_{0.7}$WS$_2$ crystals (0.1 g) were then chemically oxidized by stirring in an acidic aqueous solution of K$_2$Cr$_2$O$_7$ (0.01 mol/L) for 1 hour at room temperature. Following a series of washing procedures with distilled water and subsequent vacuum drying, the 2M-WS$_2$ crystals were obtained.

**STM measurements**

*In-situ* STM measurements were carried out in two low-temperature STM systems (Unisoku 1200 and 1300). A polycrystalline PtIr STM tip was used and calibrated using Ag island. STS data were taken by standard lock-in method. The feedback loop is disrupted during data acquisition and the frequency of oscillation signal is 811.0 Hz. $WS_2$ single crystal were cleaved at 77 K under ultra-high vacuum (with the pressure < $2\times10^{-10}$ Torr) and then transferred to STM head at 4 K.

**Data availability**

All study data are included in the article and/or *SI Appendix*.

**Acknowledgments**

We thank Y.Y. Wang, S. Kivelson, Z.Y. Weng, W.H. Duan, H. Yao, Q.H. Wang, Z.Y. Lu, K. Liu and H. Yuan for helpful discussions. The experimental work was supported by the National Natural Science Foundation of China (Grants No. 92365201, No. 52388201, No. 12234011, and No. 12374053), the National Key R&D Program of China (Grants No. 2022YFA1403100, No. 2024YFA1409100), the Science and Technology Commission of Shanghai Municipality (Grant No.24LZ1401000), Shanghai Rising-Star Program (23QA1410700), and the Innovation Program for Quantum Science and Technology (2021ZD0302400).

**Figures and Tables**

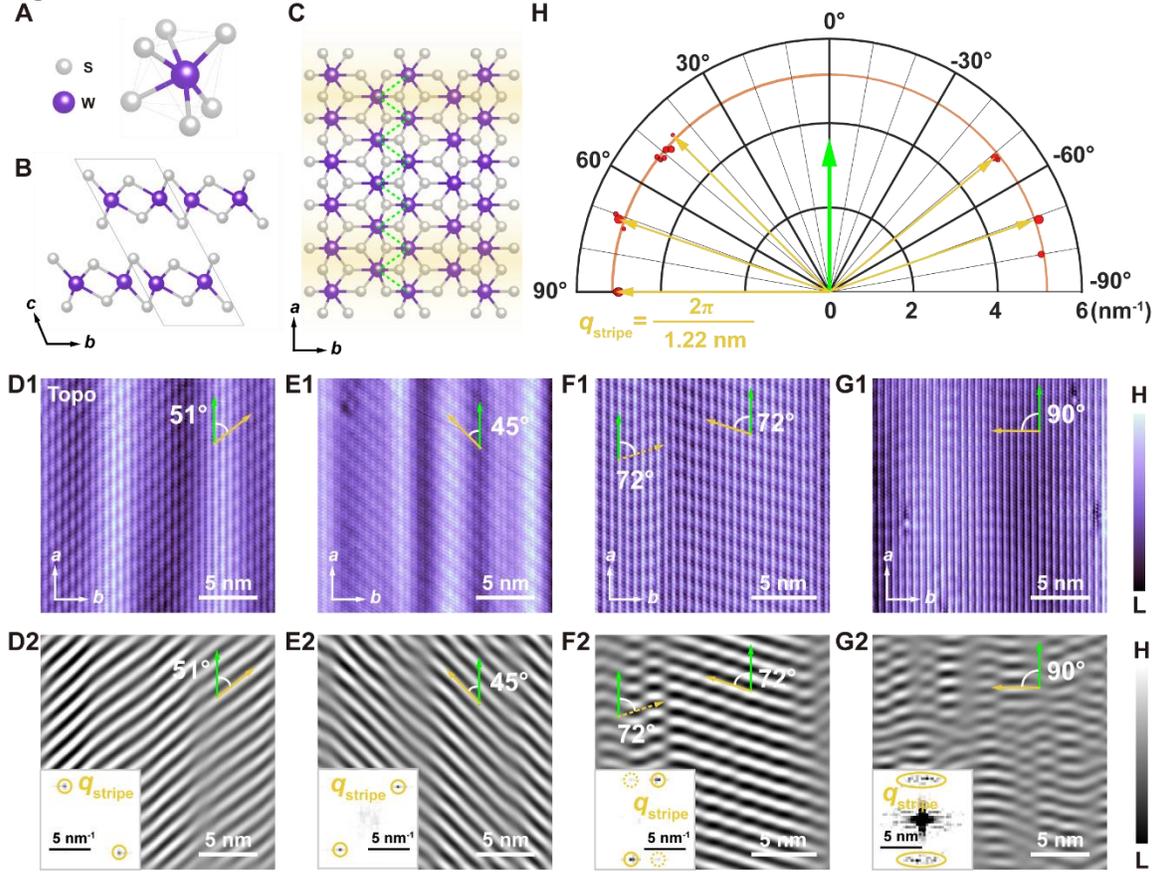

**Fig. 1.** Stripe-like charge modulations with multiple orientations. (*A-C*) The atomic structure of 2M-$WS_2$. (*A*) Schematic of $[WS_6]^{8-}$ structure. (*B*) Side view of 2M-$WS_2$. The adjacent monolayers stack along the *c*-direction through a translation operation. (*C*) Top view of 2M-$WS_2$. Peierls distortion along the *b*-direction leads to a zigzag structure along the *a*-direction denoted by green dashed lines. The yellow regions indicate the incommensurate stripe modulation perpendicular to the zigzag structure. (*D1-G1*) Atomically resolved STM topographic images of the stripe modulations along different directions (*D1*: 20 nm × 20 nm, set point, $V_s$ = 5 mV, $I_t$ = 300 pA; *E1*: 20 nm × 20 nm, set point, $V_s$ = 50 mV, $I_t$ = 100 pA; *F1*: 18 nm × 18 nm, set point, $V_s$ = 5 mV, $I_t$ = 600 pA; *G1*: 20 nm × 20 nm, set point, $V_s$ = 10 mV, $I_t$ = 1 nA). The yellow and green arrows denote the directions of the stripe modulation and the zigzag structure, respectively. The angles between the stripes and the zigzag chains can be 51°, 45°, 72° and 90° as shown in each figure. (*D2-G2*) Inverse fast Fourier transformation results of the stripes-related $q_{stripe}$ calculated from (*D1-G1*). The stripe modulations are clearly displayed. Insets: Fast Fourier transformation results of (*D1-G1*), respectively. The yellow circles denote the $q_{stripe}$ corresponding to the stripe modulations in real space. The $q_{stripe}$ shows symmetric point-like feature for 51°, 45° and 72°, while it is a line-shaped feature for 90° (24). (*H*) The statistics of the $q_{stripe}$ in different regions. The green arrow denotes the direction of the zigzag structure. The yellow arrows denote the $q_{stripe}$ in (*D-G*), respectively. The brown line indicates the average value of the $q_{stripe}$. The data presented here were collected at 4.3 K.

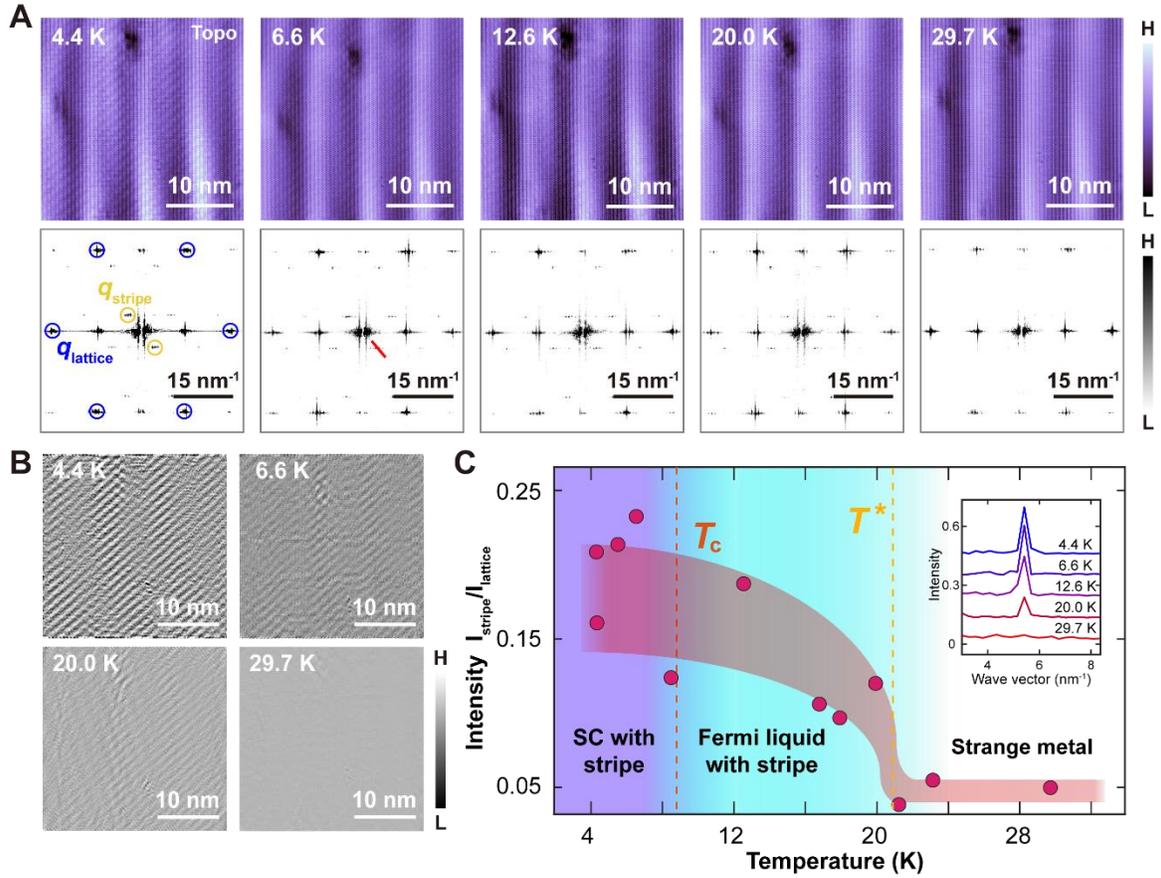

**Fig. 2.** Stripe-like charge modulation developed in the vicinity of a strange metal phase in 2M-WS$_2$. (*A*) Atomically resolved STM topographic images of the stripe modulations at different temperatures (30 nm × 30 nm; set point, $V_s$ = 5 mV, $I_t$ = 500 pA) and Fast Fourier transformation result, respectively. (*B*) Inverse fast Fourier transformation results of the signal obtained by subtracting the Moire and lattice signals from (*A*), in which the evolution of stripes are more obvious. (*C*) Temperature dependence of the wavevector $q_{stripe}$ intensity, normalized by the wavevector $q_{lattice}$ (31). The wide red curve is a guide to the eye. The orange dotted line denotes the superconducting transition temperature $T_c$ of 8.8 K. The yellow dotted line denotes the transition temperature of the charge order $T^*$ of 21 K. Inset: FFT line cuts along the red line marked in the second FFT result of (*A*) for different temperatures.

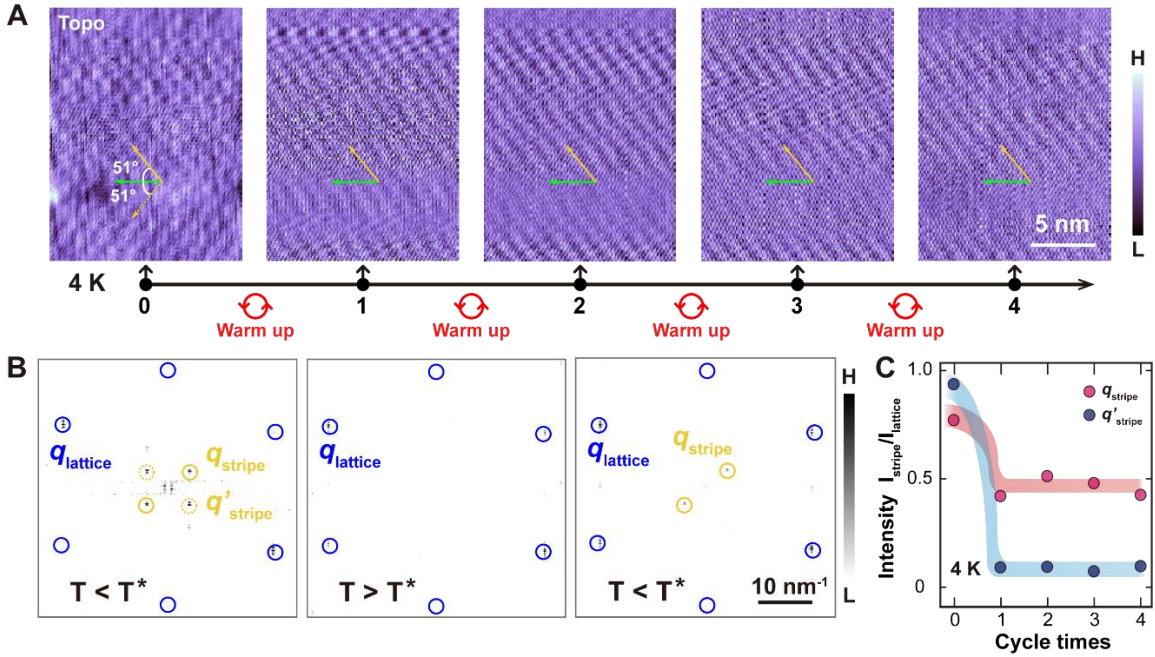

**Fig. 3.** Temperature dependent relaxation of the stripe-like charge modulation. (*A*) Topographic images of stripe modulations at 4 K under different annealing cycles (15 nm × 20 nm, for each image from left to right, the set points are respectively: $V_s$ = 10 mV, $I_t$ = 100 pA; $V_s$ = 5 mV, $I_t$ = 500 pA; $V_s$ = 5 mV, $I_t$ = 650 pA; $V_s$ = 5 mV, $I_t$ = 350 pA; $V_s$ = 5 mV, $I_t$ = 340 pA). The yellow and green arrows denote the directions of the stripe modulation and the zigzag structure, respectively. The stripes denoted by the dotted yellow arrow disappear after the first annealing cycle. The numbers on the axis represent the times of annealing cycles. (*B*) Fast Fourier transformation results of topographic images of 2M-$WS_2$ taken at different temperatures. The last image is taken at 4 K after four annealing cycles. The wavevector $q_{stripe}$ and $q_{lattice}$ correspond to the stripe modulations and the lattice, respectively. The wavevector $q'_{stripe}$ corresponds to the stripes denoted by the dotted yellow arrows in (*A*). (*C*) Cycle times dependence of the intensity of the wavevector $q_{stripe}$ and $q'_{stripe}$, normalized by the wavevector $q_{lattice}$. The wide red and blue curves are guides to the eye.

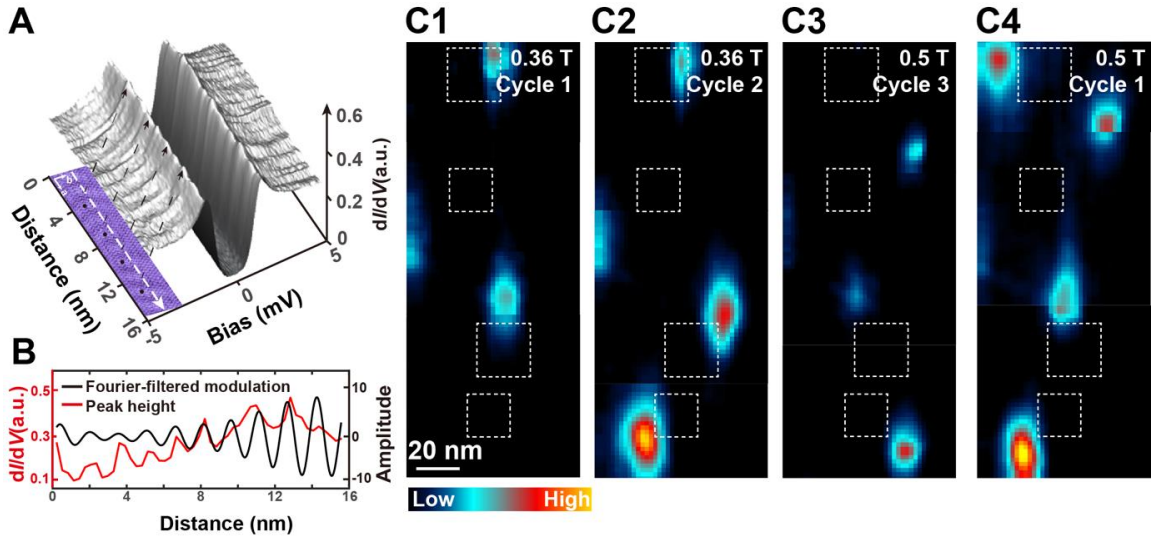

**Fig. 4.** Weak vortex pinning by the stripe-like charge modulations. (*A*) A series of d*I*/d*V* spectra taken along the white arrow in bottom inset area (Setpoint: $V_s$ = -5 mV, $I_t$ = 400 pA). Bottom inset: STM topography of stripes in 2M-$WS_2$ (3.2 nm × 15.4 nm; set point, $V_s$ = -5 mV, $I_t$ = 400 pA). The dotted black arrows link the selected peaks of the stripe patterns and the spectra obtained at these positions. (*B*) Superconducting coherence peak modulated by the stripe patterns. The red line shows the heights of the coherence peaks extracted from the spectra in (*A*). The black line is the corresponding spatial stripe modulation. (*C1-C4*) Zero-bias conductance maps under different magnetic field applying cycles (80 nm × 200 nm; set point, $V_s$ = -4 mV, $I_t$ = 400 pA). A cycle refers to a process that the magnetic field is decreased to zero and then increased to a certain value. The dashed boxes indicate the stripe-order regions. Under the same magnetic field across different cycles, the distribution of vortices varies, indicating that the pinning effect of stripes on vortices is weak. The details are shown in *SI Appendix*, Note 5. The data presented here were collected at 0.4 K.

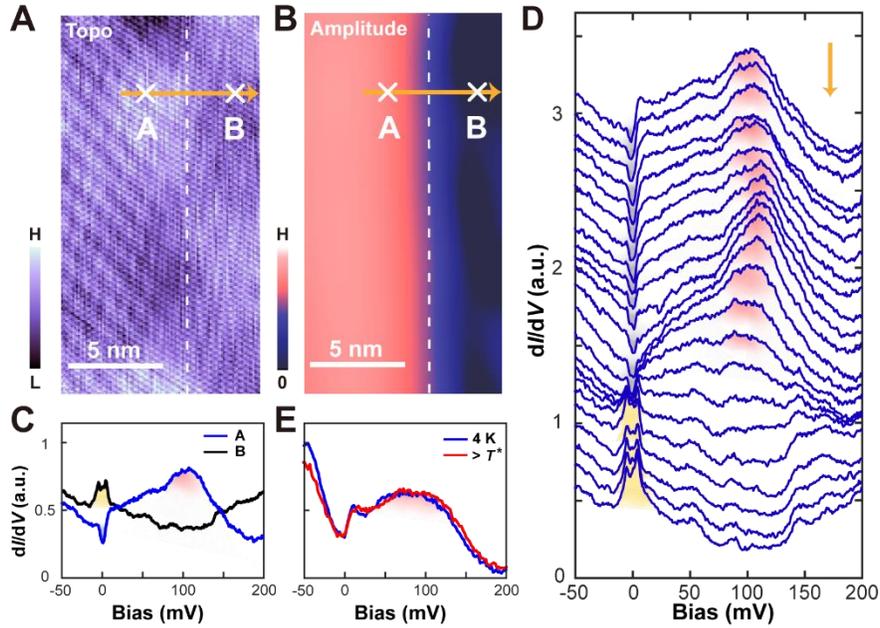

**Fig. 5.** d$I$/d$V$ spectra on the regions with and without stripes. (*A*) Topographic image of 2M-WS$_2$ (10.5 nm × 20 nm; set point, $V_s$ = 50 mV, $I_t$ = 100 pA), in which stripe and no-stripe regions are displayed. The white dashed line denoted the boundary between the stripe and no-stripe regions. (*B*) Amplitude map of the wavevector $q_{stripe}$, in the same area of (*A*). The details are shown in *SI Appendix*, Fig. S10. (*C*) d$I$/d$V$ spectra (Set point: $V_s$ = 200 mV, $I_t$ = 250 pA) taken at the points denoted by the white crosses in (*A*). The yellow shade denotes the DOS increasing around $E_F$ in the spectrum taken at the no-stripe region (point B). The red shade denotes the DOS increasing around 100 meV in the spectrum taken at the stripe regions (point A). The gray shade denotes the superconducting gaps. (*D*) A Series of d$I$/d$V$ spectra (Set point: $V_s$ = 200 mV, $I_t$ = 250 pA) taken along the orange arrow in (*A*). The yellow, red and gray shades have the same meaning as in (*C*). (*E*) d$I$/d$V$ spectra (Set point: $V_s$ = -500 mV, $I_t$ = 800 pA) of another stripe region taken below and above $T^*$, respectively. The 100 meV peak-features are unchanged. Note that the superconducting gap here is not visible in the large-energy-range spectrum, due to an even larger background dip-feature (location dependent) near $E_F$. The data presented here were collected at 4.3 K.